\RequirePackage{fix-cm}
\documentclass{svjour3}                     % onecolumn (standard format)
\smartqed  % flush right qed marks, e.g. at end of proof
\usepackage{graphicx}
\usepackage{natbib}
\usepackage{amsmath}
\journalname{Experimental Astronomy}
\begin{document}

\title{SPIRE Point Source Photometry.
}
\subtitle{within the Herschel\footnote{{\it Herschel} is an ESA space observatory with science
instruments provided by European-led Principal Investigator consortia and with 
important participation from NASA.} Interactive Processing Environment (HIPE)\\ }


\author{Chris Pearson         \and
        Tanya Lim         \and
         Chris North         \and
        George Bendo    \and
	Luca Conversi     \and
	Darren Dowell  \and
	Matt Griffin  \and
	Terry Jin   \and
	Nicolas Laporte \and
	Andreas Papageorgiou \and
	Bernhard Schulz \and
	Dave Shupe \and
	Anthony J. Smith \and
	Kevin Xu
}


\institute{C.P. Pearson \at
              RAL Space, STFC Rutherford Appleton Laboratory, Didcot, Oxon, OX11 0QX, UK \\
              The Open University, Milton Keynes MK7 6AA, UK \\
              Tel.: +44 1235 44 5047,  \email{chris.pearson@stfc.ac.uk}   \\
           \and
           T. Lim \at
              RAL Space, STFC Rutherford Appleton Laboratory, Didcot, Oxon, OX11 0QX, UK \\
                         \and
           C.E. North, M.J. Griffin, A. Papageorgiou  \at
              School of Physics and Astronomy, Cardiff University, The Parade, Cardiff CF24 3AA, UK \\
           \and
            G.J. Bendo \at
              UK ALMA Regional Centre Node, Jodrell Bank Centre for Astrophysics, School of Physics and Astronomy, University of Manchester, Oxford Road, Manchester M13 9PL, United Kingdom \\     
              \and
              L. Conversi \at 
              Herschel Science Centre, ESAC, ESA, Villanueva de la Ca÷nada, 28691 Madrid, Spain \\
              \and
              C. D. Dowell \at 
              NASA Jet Propulsion Laboratory, 4800 Oak Grove Drive, Pasadena, CA 91109, USA \\
              \and
              T.Jin \at
             Department of Physics and Astronomy, University College London, London WC1E 6BT, UK \\
             \and
             N. Laporte \at
              Instituto de Astrofisica de Canarias, Calle Via Lactea, 38205 - La Laguna, Spain \\
                            \and
              D.L. Shupe, B. Schulz, C.K. Xu \at 
             NASA Herschel Science Centre, IPAC, 770 South Wilson Avenue, Pasadena, CA 91125, USA \\
             \and 
             A.J.Smith \at
             Astronomy Centre,  University of Sussex, Brighton, BN1 9QH, UK  \\      
             Bluesky Spectroscopy, Lethbridge, Canada \\       
           }

\date{Received: 30th June 2013 / Accepted: date}

\maketitle

% *************************************************************************************************
% ***********                           Section:Abstract             *********************************
% *************************************************************************************************
\begin{abstract}
The different algorithms appropriate for  point source photometry  on data from the SPIRE instrument on-board the {\it Herschel} Space Observatory, within the Herschel Interactive Processing Environment (HIPE) are compared. Point source photometry of a large ensemble of standard calibration stars and dark sky observations is carried out using the 4 major methods within HIPE: SUSSEXtractor, DAOphot, the SPIRE Timeline Fitter and simple Aperture Photometry. Colour corrections and  effective beam areas as a function of the assumed source spectral index are also included to produce a large number of photometric measurements per individual target, in each of the 3 SPIRE bands (250, 350, 500$\mu$m), to examine both the accuracy and repeatability  of each of the 4 algorithms. It is concluded that for flux densities down to the level of 30mJy that the SPIRE Timeline Fitter is the method of choice. However, at least in the 250 and 350$\mu$m bands, all 4 methods provide photometric repeatability  better than a few percent down to at approximately 100mJy. The DAOphot method appears in many cases to have a systematic offset of $\sim$8$\%$ in all SPIRE bands which may be indicative of a sub-optimal aperture correction. In general,  aperture photometry is the least reliable method, i.e. largest scatter between observations, especially in the longest wavelength band. At the faintest fluxes, $<$30mJy, SUSSEXtractor or DAOphot  provide a better alternative to the Timeline Fitter.
\end{abstract}

% *************************************************************************************************
% ***********                      Section: Introduction            *********************************
% *************************************************************************************************
\section{Introduction}
\label{sec:introduction}
The SPIRE instrument \citep{griffin10} on board the {\it Herschel} Space Observatory  \citep{pilbratt10} with its 3 photometric bands at 250, 350 \& 500$\mu$m (referred to as the PSW, PMW \& PLW bands respectively) has allowed, for the first time, the accumulation of large sample point source data sets at submillimetre wavelengths for both galactic and extragalactic programmes (e.g. \citet{eales10}, \citet{molinari10}, \citet{oliver12}). Standard pipeline reduction of SPIRE mapping observations is made within the Herschel Common Science System  {\it Herschel Interactive Processing Environment} (HIPE \citet{ott10}) with SPIRE maps often requiring little or no reprocessing compared to the archive products, due to the extremely stable nature of the SPIRE instrument. Therefore, first order analysis of the maps produced by SPIRE is the extraction and subsequent photometry of detected point sources. Although there exist a vast menagerie of candidate algorithms for photometry (e.g. \citet{bertin96}, \citet{diolaiti00}),  in this work, we present the photometry algorithms available within HIPE itself. The aim of work is to quantify the reliability of each photometry method relative to each other over a large range of observations both in flux density and time during the mission. Here we present the results for measurements using a set of SPIRE standard calibration stars and dark field observations. A full treatment of the photometry of the SPIRE calibration stars and asteroids is given in Lim et al. (2013, in preparation), including a comparison with contemptrary stellar models. In Section~\ref{sec:spiremaps}, we described the SPIRE map calibration philosophy, in Section~\ref{sec:photometrymethods} the various photometry algorithms available with HIPE are introduced. The photometry tests and results for the calibration stars and dark skies are explained in Section~\ref{sec:photometrytests} and the summary and conclusions is given in Section~\ref{sec:summary}.

% *************************************************************************************************
% ***********           Section: Map Calibration                 *********************************
% *************************************************************************************************
\section{SPIRE Map Calibration}
\label{sec:spiremaps}
SPIRE maps are constructed from the individual timeline data. As a detector is scanned directly over a point source, the peak deflection of the signal timeline equals the brightness of the source. The SPIRE calibration applies to the timelines (referred to as the Level 1 products) which are calibrated in Jy/beam with the peak value corresponding to the source flux density (note that mapmaking will {\it lower} these peaks). The SPIRE photometer pipeline (\citet{dowell10}, Pearson et al., in preparation)  produces monochromatic in-beam flux densities (Jy/beam) at frequencies corresponding to 250, 350 and 500$\mu$m, under the assumption that the source is point-like with a flat spectrum $\nu F_{\nu} = $ constant.
The SPIRE photometer maps (referred to as Level-2)  are also calibrated in terms of in-beam flux density (Jy/beam) rather than surface brightness (Jy/pixel, MJy/sr, etc).
SPIRE photometer calibration is based on measurements of Neptune, both for the flux density and model beams and are described in detail in \citet{bendo13}. The current calibration accuracies are $\sim$1.5$\%$ for instrumental uncertainties and $\sim$4$\%$ uncertainties in the model flux density of Neptune. 

The SPIRE beams are measured on Neptune, which has a particular spectral index, $F_{\nu} \propto \nu^{\alpha}$, $\alpha_{Neptune}$= 1.29, 1.42, 1.47 for the PSW, PMW, PLW bands respectively (\citet{moreno98}, \citet{moreno12}, \citet{bendo13}). Moreover, the SPIRE data reduction pipeline assumes an $\alpha_{pipeline}$=-1. Therefore  corrections must  be made for sources with a different spectrum. This involves taking into account the frequency-dependence of the beam with a Full Width Half Maximum (FWHM)  varying with frequency as $\nu^{-0.85}$ (see \citet{griffin13} for a detailed derivation). Thus the monochromatic beam solid angle, $\Omega_{\nu}(\nu)$ varies as Equation~\ref{eqn:beam} and the effective beam area, $\Omega_{eff}(\alpha)$, integrated over the instrument relative spectral response function (RSRF), $R_{\nu}$, and source spectrum, $\nu^{\alpha}$, vary as in  Equation~\ref{eqn:effbeam}, where $\nu_{eff}$ is the frequency at which the monochromatic beam solid angle equals the beam as measured on Neptune, i.e. $\Omega_{\nu}(\nu_{eff})$ = $\Omega_{eff}(\alpha_{Neptune})$ = $\Omega_{Neptune}$. This effective frequency is necessary for converting a map between flux density (Jy/beam) and surface brightness (Jy/pixel) units. The pipeline beam values $\Omega_{eff}(\alpha)$=-1 are listed in Table~\ref{Tab:PhotometryParameters}.

% ************************* BEGIN EQUATION *************************
\begin{equation}\label{eqn:beam}
\Omega_{\nu}(\nu) = \Omega_{Neptune} \left({\nu\over{\nu_{eff}}}\right)^{-1.7}
\end{equation}
% ************************* END EQUATION *************************

% ************************* BEGIN EQUATION *************************
\begin{equation}\label{eqn:effbeam}
%\Omega_{eff}(\alpha) = {\int R_{\nu}\nu^{\alpha}\Omega_{\nu}(\nu)\rmn{d}\nu \over{ 1  }}
\Omega_{eff}(\alpha) = {\int R_{\nu}\nu^{\alpha}\Omega_{\nu}(\nu) d\nu \over{\int R_{\nu}\nu^{\alpha}d\nu}} =  {\int R_{\nu}\nu^{\alpha} \left({\nu\over{\nu_{eff}}}\right)^{-1.7} d\nu \over{\int R_{\nu}\nu^{\alpha}d\nu}}
\end{equation}
% ************************* END EQUATION *************************

In addition to the effective beam areas, point source colour corrections ($K_{colP}(\alpha)$) are also required to give the flux density of a point source with a different spectral index ($\alpha$), compared to the pipeline assumption of $\alpha$=-1 (see Equation~\ref{eqn:colourcorr}, also \citet{griffin13}). Where  $\eta_{\nu}$ is the aperture efficiency of the telescope and $\nu_{o}$ is the monochromatic frequency of the SPIRE bands at 250, 350, 500$\mu$m respectively. 

% ************************* BEGIN EQUATION *************************
\begin{equation}\label{eqn:colourcorr}
%\Omega_{eff}(\alpha) = {\int R_{\nu}\nu^{\alpha}\Omega_{\nu}(\nu)\rmn{d}\nu \over{ 1  }}
K_{colP}(\alpha) =  {\int R_{\nu}  \left({\nu\over{\nu_{o}}}\right)^{-1} \eta_{\nu} d\nu \over{\int R_{\nu}  \left({\nu\over{\nu_{o}}}\right)^{\alpha} \eta_{\nu} d\nu}}
\end{equation}
% ************************* END EQUATION *************************

% *************************************************************************************************
% ***********           Section: Photometry Algorithms             ***************************
% *************************************************************************************************
\section{Photometry Methods within HIPE}
\label{sec:photometrymethods}
There are various methods for photometry of point sources within the HIPE system, with some algorithms facilitating both source extraction and photometry and others just providing photometry alone. In this work we discuss the 4 main options available for SPIRE point source photometry, the 2 source extractors; SUSSEXtractor and DAOphot and the 2 photometry algorithms; Aperture Photometry and the Timeline Fitter. A brief description of each algorithm is given below and the parameters used for this work are shown in Table~\ref{Tab:PhotometryParameters}. In Table 1, the effective beams and FWHM are those derived from observations of Neptune \citep{griffin13}. The adopted values for the apertures and background annuli are described individually for each photometry method below.

\subsection{SUSSEXtractor}
\label{sec:SUSSEXtractor}
The SUSSEXtractor algorithm is described in detail in \citet{savage07} and employs a Bayesian approach, modelling both the source and the empty sky at each potential source position. For each model, parameters and likelihoods are estimated, and then the models are compared in order to determine whether a source is present at that location. The image is smoothed with a convolution kernel, derived from the point response function (PRF) and the resulting smoothed image is searched for local maxima (peaks in the image). A local maximum is a pixel which is higher than all of its neighbours within a pixel distance defined by a {\it pixelRegion} parameter. The pixel region value adopted is the default value of 1.5.  The intensity in the smoothed image at the position of a point source is taken as the estimate of the source flux density. Only those sources with a threshold above some specified detection threshold are then accepted as detections.
SUSSEXtractor requires an image map and noise map as input, along with parameters for the Full Width Half Maximum (FWHM) and detection threshold. No aperture correction is required (see Table~\ref{Tab:PhotometryParameters}).
The output from SUSSEXtractor is a {\it Source List Product} containing the position of the source in pixel coordinates and R.A. \& Dec, the measured source flux density with associated errors, the measured background flux density with associated errors and the threshold at which a local maximum was detected (this is the  log(Bayes factor), i.e. the  difference in log evidence between the source model and the background model, similar to a likelihood ratio). 

\subsection{DAOphot}
\label{sec:DAOphot}
The HIPE implementation of DAOphot incorporates the classic DAOPHOT algorithm \citep{stetson87}, using the FIND and APER procedures in the IDL\footnote{Interactive  data Language: http://www.exelisvis.com/ProductsServices/IDL.aspx} Astronomy User's Library \footnote{http://idlastro.gsfc.nasa.gov/contents.html\#C2}. The image is smoothed with a DAOPHOT convolution kernel, scaled from the Gaussian beam profile.  The convolved image is searched for local maxima (peaks in the image) with similar criteria to SUSSEXTractor. The Photometry is carried out using circular aperture photometry using a source aperture (default = 1 $\times$ FWHM from the original IDL implementation) and a background annulus (default = 1.25 - 3 $\times$ FWHM), with the source flux calculated as the sum of the pixels in the source aperture minus  the background value subtracted from each pixel. Note that the aperture photometry requires the image to be in units of surface brightness (e.g. Jy/pixel), therefore an internal conversion is made using the supplied beam area. DAOphot only requires an image map as input and the beam area, FWHM and detection threshold as parameters (see Table~\ref{Tab:PhotometryParameters}). The HIPE DAOphot algorithm also calculates an aperture correction automatically. Aperture corrections are calculated by performing internal aperture photometry on an image of the model (Gaussian) beam assuming a model true flux of 1 Jy.
The  output from DAOphot is also a {\it Source List Product} containing the position of the source in pixel coordinates and R.A. \& Dec, the measured source flux density with associated errors, the measured background flux density with associated errors and DAOphot roundness/sharpness parameters.

\subsection{SPIRE Timeline Fitter}
\label{sec:TimelineFitter}
The SPIRE Timeline Fitter (see \citet{bendo13} for a detailed description of the methodology) does not work with image maps, instead it uses a Levenberg-Marquardt algorithm to fit two dimensional elliptical (or circular) Gaussian functions to the 2-D timeline data, projected on to a zero-footprint map, thus avoiding smearing effects related to pixelisation or drizzling associated with the map making process. Fitting to the timeline data enables a more precise measurement of the source peaks which for timelines calibrated in Jy/beam will be the source flux density. The  SPIRE Timeline Fitter requires baseline-subtracted (destriped) timelines and an initial source position which is refined by the fitting process. In addition, the Timeline Fitter requires a search radius (in arcsec) that includes the peak of the source as a parameter (see Table~\ref{Tab:PhotometryParameters}). A Gaussian is fit to the source and background as defined by a background annulus. The default values for the aperture and background annulus are derived from observations of Neptune and Gamma Draconis \citep{bendo13}. The  output from the Timeline Fitter is also a {\it Source List Product} containing the position of the source in  R.A. \& Dec with associated errors, the measured flux density of the source and background  with associated errors and the major and minor axis of the fitted Gaussian with associated $\chi ^2$ fit.

\subsection{Aperture Photometry}
\label{sec:Aperture}
Simple annular sky aperture photometry is supported within HIPE as a Java task. The input for the Aperture Photometry is an image with units of surface brightness (specifically in Jy/pixel), therefore the input image map (in Jy/beam) must first be divided by the beam area.  The Photometry is carried out by defining a source aperture  and a background annulus. The  values adopted in  Table~\ref{Tab:PhotometryParameters} are the HIPE default values however for point sources, varying the parameters negligibly  effects the results. The source flux density is then calculated as the sum of the pixels in the source aperture minus  the background value subtracted from each pixel. In addition to the source aperture and background annuli, the Aperture Photometry task also requires the source position (centre of the source aperture) in RA \& Dec (see Table~\ref{Tab:PhotometryParameters}). the output from the Aperture Photometry is a HIPE {\it product}, including the source flux estimate, background estimate and curve of growth. Note that the errors from the HIPE Aperture Photometry task are currently not reliable and in addition no aperture correction is applied but is required.

%   *************************  BEGIN TABLE  *************************
\begin{table*}
\caption{Parameters used for the four different photometry algorithms within HIPE (beam areas are the default pipeline values assuming a source spectrum $F_{\nu}\propto \nu^{\alpha}$, $\alpha$ = -1). Where appropriate (e.g. effective beam area) parameters are listed consecutively for the SPIRE PSW (250$\mu$m), PMW (350$\mu$m), PLW (500$\mu$m)  bands respectively.}
\label{Tab:PhotometryParameters}       % Give a unique label
\begin{tabular}{lllll}
\hline\noalign{\smallskip}
Parameter & SUSSEXtractor & DAOphot  & Timeline & Aperture\\
\noalign{\smallskip}\hline\noalign{\smallskip}
Eff. Beam Area (arcsec$^2$) & - & 465, 822, 1769 & - & -\\
FWHM ($^{\prime\prime}$) & 17.6,23.9,35.2 & 17.6,23.9,35.2 & - & -\\
Aperture  ($^{\prime\prime}$) & - & 1$\times$FWHM & 22,30,42 & 22,30,42\\
Background  ($^{\prime\prime}$) & - & 1.25-3$\times$FWHM & 300-350 & 60-90\\
Aperture Correction & - & automatic & - & 1.28, 1.196, 1.26 \\
\noalign{\smallskip}\hline
\end{tabular}
\end{table*}
%   *************************  END TABLE  *************************

\subsection{Procedure for photometry within HIPE}
\label{sec:HIPEphotometry}
The general procedure for processing map observations for photometry depends on the photometry method that will be employed. Broadly speaking the photometry can be divided into two types; those methods that fit the source flux either in the map or the timelines (SUSSEXtractor, Timeline Fitter) and those that add up the flux contained within the map pixels associated with the source (DAOphot, Aperture Photometry).

\begin{figure*}
\centering
  \includegraphics[width=0.75\textwidth]{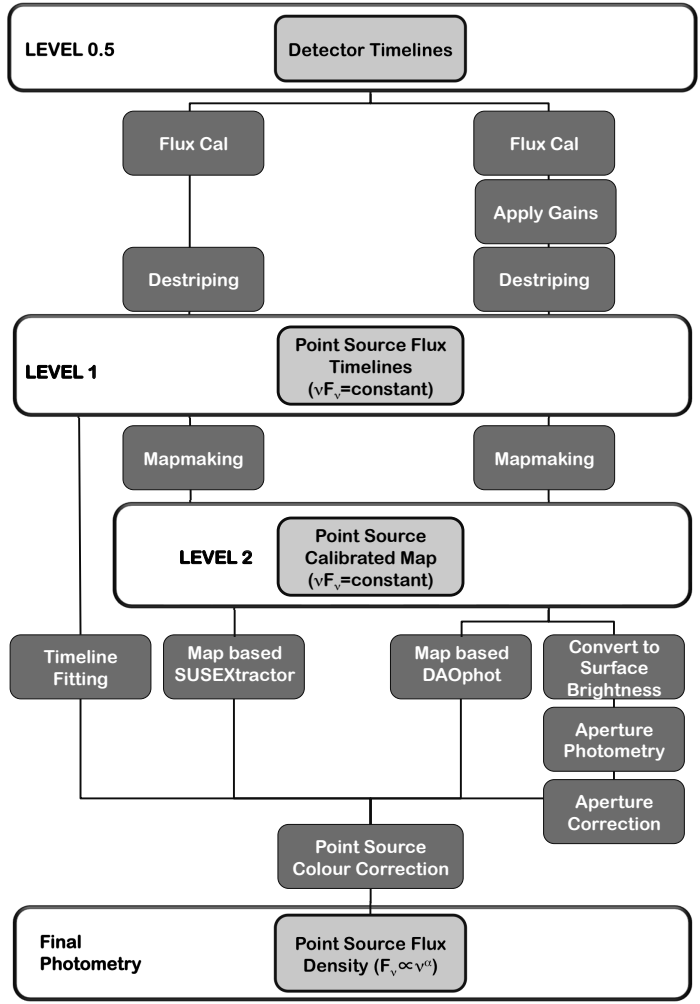}
\caption{Algorithmic flow diagram for carrying out point source photometry for different photometry tasks within the SPIRE calibration framework within HIPE.}
\label{fig:photometryCalFlowchart}       
\end{figure*}

In Figure~\ref{fig:photometryCalFlowchart} the algorithmic pipeline is shown. The starting point is the Level 0.5 timeline data (as voltage time streams) which are then processed through the Level 0.5 - Level 1 pipeline to remove instrumental effects such as crosstalk, responsivity corrections, and other artefacts. {\it Flux Calibration} is made on the timelines following ~\citet{bendo13}. For optimal photometry using aperture methods, an extra processing step, referred to as the {\it Apply Relative Gains} task, is optionally desirable. Since not all detector beam-profiles  have the same width there is a variation of the beam profiles across the detector arrays, the  {\it Apply Relative Gains} module takes into account  these  beam shape variations to correct for the differences in the ratio of integral verses the peak of each beam profile. This is important since although the timelines are calibrated in Jy/beam with the peak giving the source flux density, aperture photometry methods effectively depend on the integral of the beam. 

The flux calibrated timelines are then destriped (Schulz et al. 2013, in preparation) to remove any drifts in the timeline data that would result in stripes in the final maps. The final Level 1 SPIRE data products are the destriped, Jy/beam flux calibrated timelines. The SPIRE Timeline Fitter works on these timelines. 

The final SPIRE Level 2 map products are then constructed using the Level 1 timelines by the mapmaking module which uses a naive mapmaking algorithm. The Naive Mapmaker simply projects the full power seen by a detector onto the nearest sky map pixel. For each detector timeline, at each time sample, the signal measurement is added to the total signal map, the square of the signal is added to the total signal squared map, and 1 is added into the coverage map. After all detector signals have been mapped, the total signal map is divided by the coverage map to produce a flux density map, and the standard deviations are calculated using the total signal, total signal squared, and coverage map. The SUSSEXtractor, DAOphot and Aperture Photometry algorithms work on these maps. Note that for the specific case of the HIPE Aperture Photometry, the maps need to be divided by the beams since the task expects a map in Jy/pixel units (DAOphot makes this conversion internally). In addition, the fluxes from the Aperture Photometry  also require  aperture corrections, given in Table~\ref{Tab:PhotometryParameters}. The final step is to apply the appropriate colour corrections for the assumed spectral index of the source to the measured fluxes.

\begin{figure*}
\centering
  \includegraphics[width=0.6\textwidth]{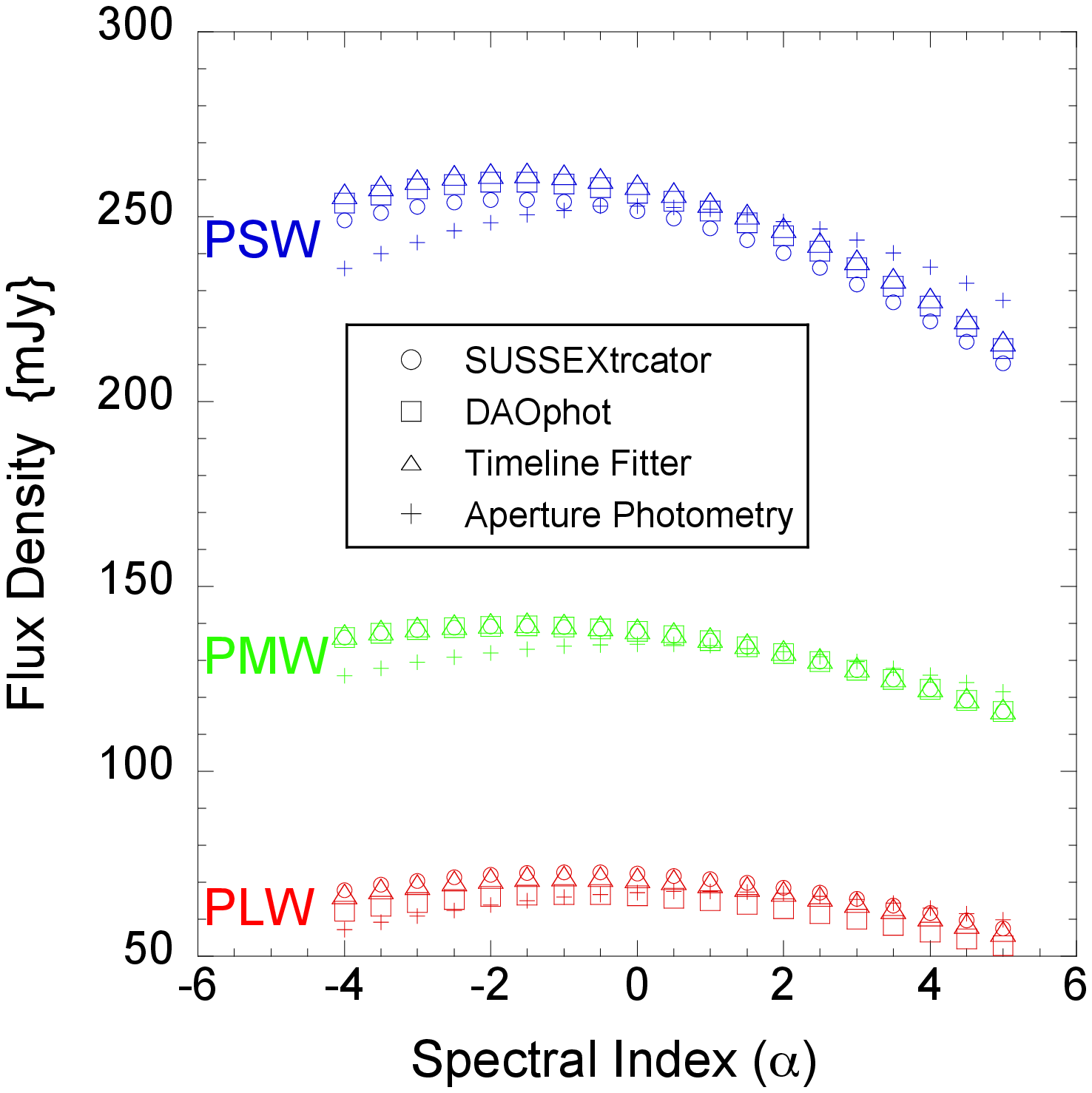}
\caption{The effect of the colour correction and beam area on the measured flux density of the SPIRE standard calibration star, Gamma Draconis. Results for all 4 photometry methods are shown for the SPIRE PSW (250$\mu$m), PMW (350$\mu$m), PLW (500$\mu$m)  bands. Note that the beam variation only affects the Aperture Photometry method.}
\label{fig:beamvariation}       
\end{figure*}

Note that the correct choice of beam and colour correction for a given spectral index ($\alpha$) will effect the final flux measured. In Figure~\ref{fig:beamvariation}, the effect of the colour correction and the effective beam area, for an observation of a SPIRE calibration star, Gamma Dra is shown for the 4 photometry methods as a function of spectral index. The figure shows that the colour correction can produce up to 20$\%$ variations in the interpreted flux density in the extreme. In addition, for the methods involving aperture photometry the beam is required to convert the map from Jy/beam units to Jy/pixel units. The Timeline Fitter and SUSSEXtractor are independent of the beam, as is DAOphot if the automatic aperture correction option is turned on.  From Figure ~\ref{fig:beamvariation}, it can be seen that the aperture photometry is also affected by the beam. Note for the calibration stars in general, the photosphere corresponds to $\alpha \sim$2 at which point all the photometry methods including the aperture photometry are in broad agreement.

%
% ************************* BEGIN FIGURE *************************
% ***************************************************************************

%
% ************************* BEGIN FIGURE *************************
% ***************************************************************************

% *************************************************************************************************
% ***********           Section: Photometry Tests             *********************************
% *************************************************************************************************
\section{Testing the Photometry Methods}   \label{sec:photometrytests}

In order to test the agreement between the various photometry algorithms available within HIPE, each method was run on an identical ensemble of SPIRE standard calibration stars and on a set of mapping observations in the SPIRE dark sky area. Note that SPIRE photometry measurements using Asteroids will be discussed in Lim et al. (2013, in preparation). All observations were processed using Version 11 of the Herschel Common Science System  {\it Herschel Interactive Processing Environment} (HIPE \citet{ott10}) using the standard user pipeline \citep{dowell10}, with default values for all tasks, utilizing the SPIRE Calibration Tree version 11.0.  The difference between HIPE 11 compared to previous versions of HIPE is the inclusion of new flux calibration products, in the calibration tree used in the pipeline processing. This updated flux calibration is based on  the new ESA4 models of Neptune \citep{moreno12}  and results in small changes of $\sim <$ few $\%$ in the measured flux of sources.

\subsection{Standard Calibration Stars}
\label{sec:standardstars}
 The standard stars used and the number of observations processed are given in Table~\ref{Tab:Standardstars}. All observations were made between SPIRE Operational Days (OD) 100-1100 corresponding to 21st August 2009 -- 17th May 2012. We assume  stellar photospheres corresponding to a spectral index of $\alpha$=2 and apply the appropriate beams and colour corrections given in \citet{griffin13} and Equations~\ref{eqn:effbeam},~\ref{eqn:colourcorr}. The beams appropriate for a spectral index, $\alpha$=2, are 445, 788 \& 1645 square arcsec for the PSW, PMW \& PLW bands respectively. The corresponding colour corrections are 0.9454, 0.9481 \& 0.9432 for the PSW, PMW \& PLW bands respectively. Stars are located in the final maps using SUSSEXtractor to provide the target position (R.A., Dec.). This position is then used as input to the other photometry methods.

%   *************************  BEGIN TABLE  *************************
\begin{table*}
\caption{SPIRE standard calibration stars used to test the photometry methods available in HIPE. R.A. and Dec are given in J2000 sexadecimal format. The number of observations analysed for each star is given in the right-most column}
\label{Tab:Standardstars}       % Give a unique label
\begin{tabular}{llll}
\hline\noalign{\smallskip}
Star & RA & Dec  & Number \\
\noalign{\smallskip}\hline\noalign{\smallskip}
Alpha Boo & 14h15m39.670s  & +19d10m56.70s & 10\\
Alpha Tau & 4h35m55.240s  & +16d30m33.50s & 11\\
Alpha Ari & 2h07m10.290s  & +23d27m46.00s & 6\\
Gamma Dra & 17h56m36.370s  & +51d29m20.00s & 84\\
Alpha Cet & 3h02m16.780s  & +4d05m23.70s & 6\\
Alpha Hya & 9h27m35.240s  & -8d39m31.00s & 7\\
HR 7557 & 19h50m47.000s  & +8d52m06.00s & 4\\
Beta Peg & 23h03m46.460s  & +28d04m58.00s & 12\\
Beta Umi & 14h50m42.330s  & +74d09m19.80s & 21\\
Gamma Cru & 12h31m09.930s  & -57d06m45.20s & 7\\
Sirrius & 6h45m08.920s  & -16d42m58.00s & 9\\
Beta Gem & 7h45m18.950s  & +28d01m34.30s & 9\\
Beta And & 1h09m43.800s  & +35d37m15.00s & 5\\
Epsilon Lep & 5h05m27.670s  & -22d22m15.70s & 5\\
Omega Cap & 20h51m49.290s  & -26d55m08.90s & 5\\
\noalign{\smallskip}\hline
\end{tabular}
\end{table*}
%   *************************  END TABLE  *************************

% ************************* BEGIN FIGURE *************************
\begin{figure}
\centering
  \includegraphics[width=0.32\textwidth]{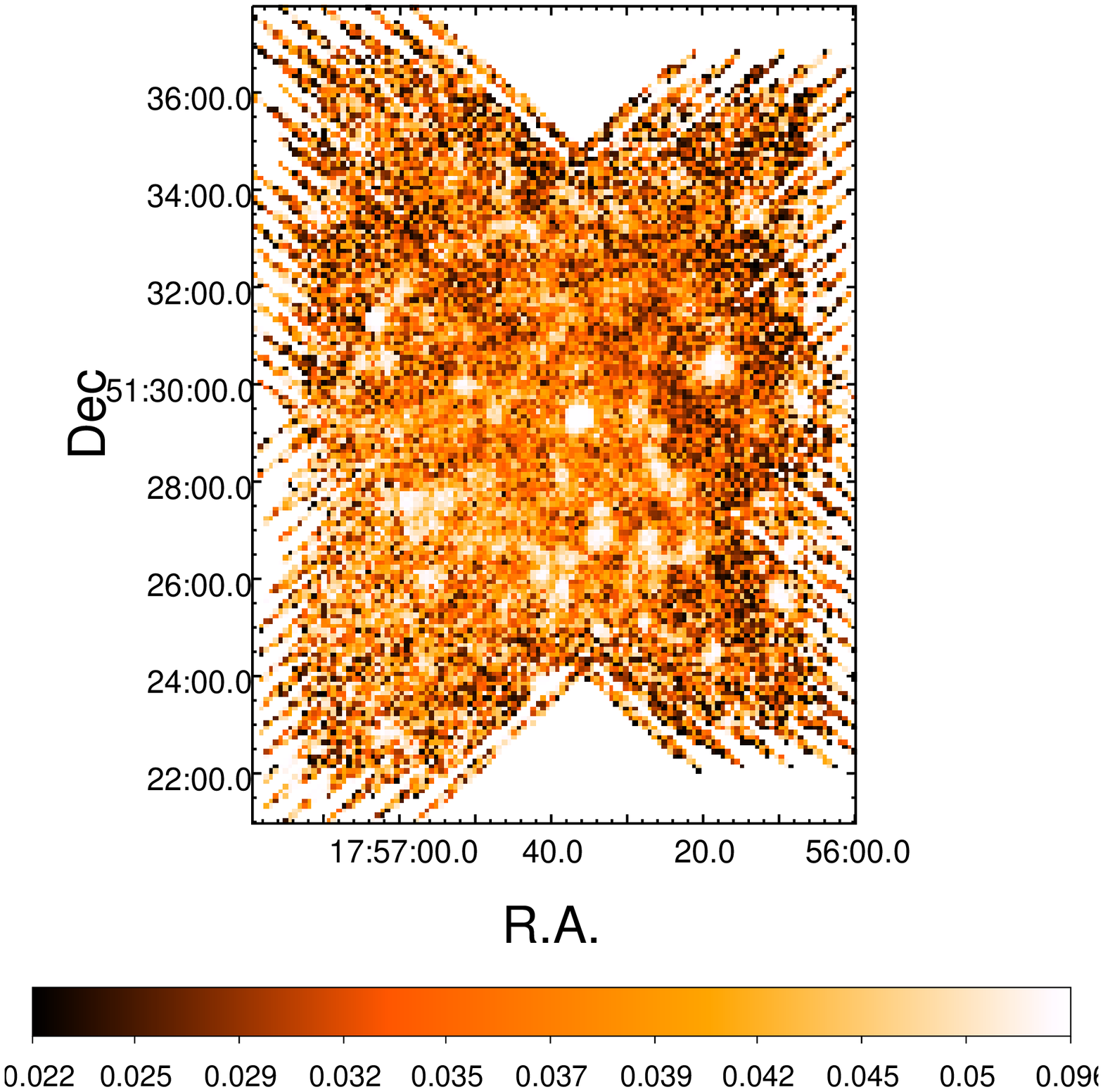}
    \includegraphics[width=0.32\textwidth]{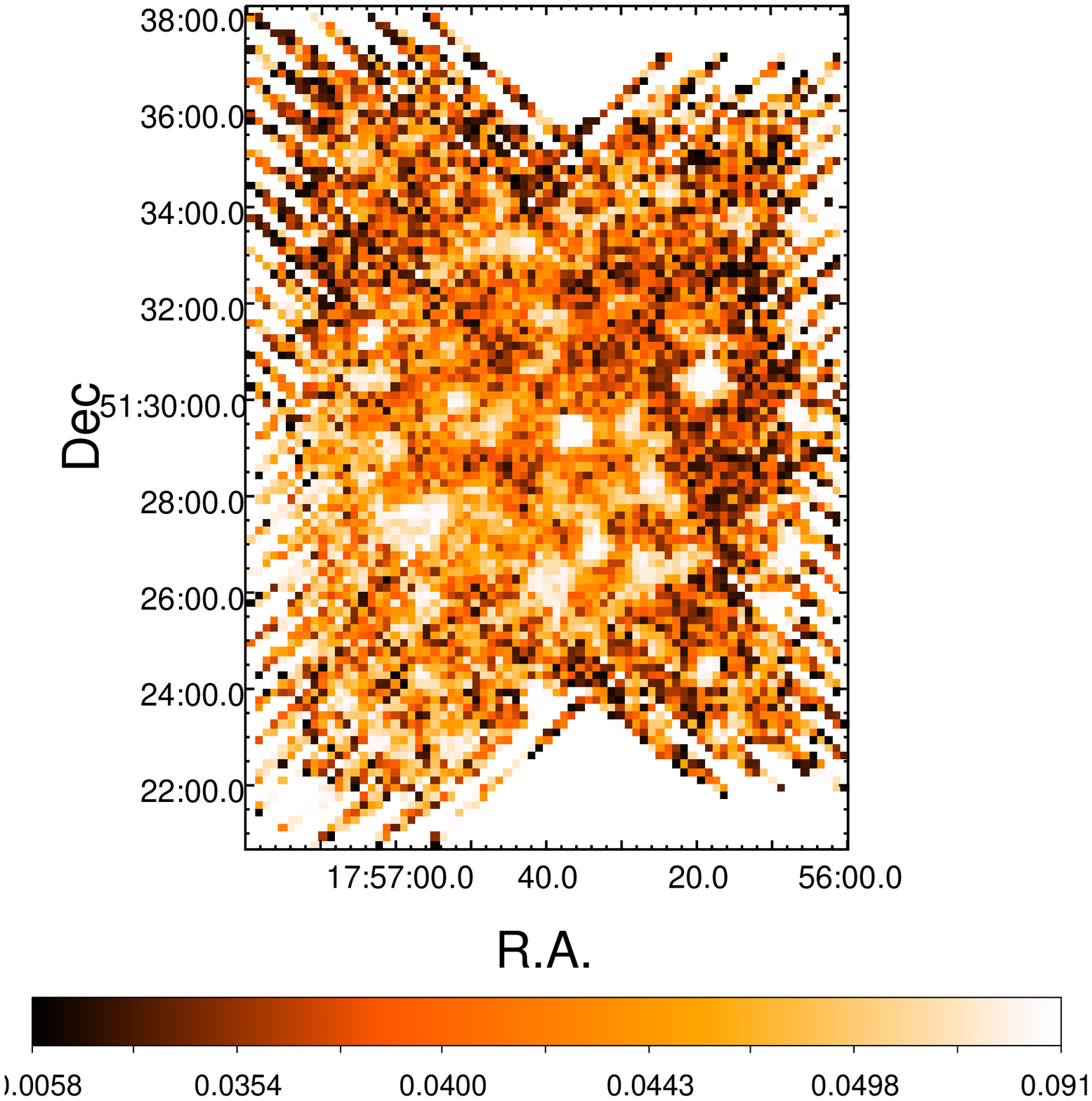}
  \includegraphics[width=0.32\textwidth]{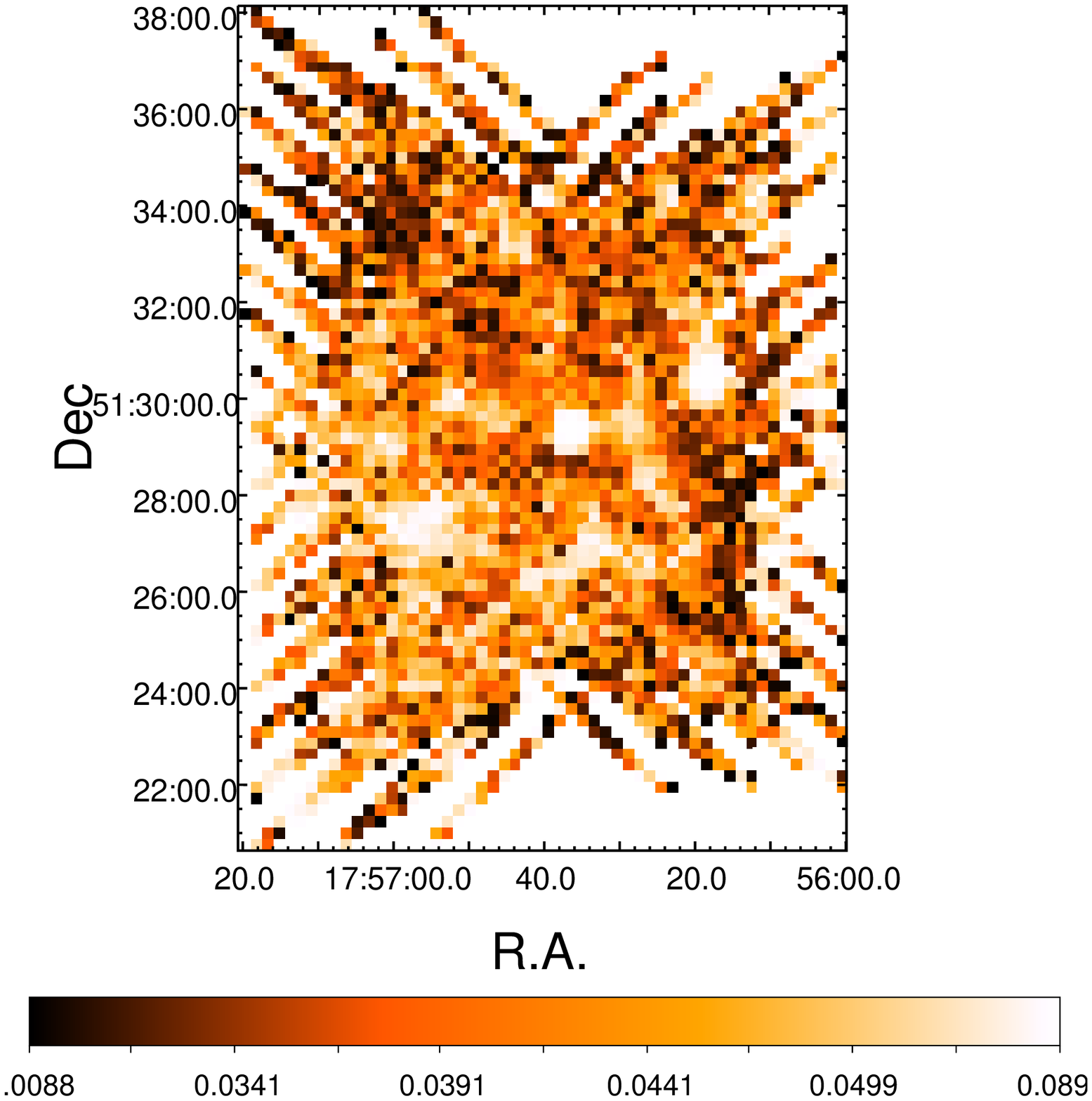}
\caption{Final SPIRE maps for the standard calibration star Gamma Dra, PSW (left), PMW (middle) and PLW (right) bands. The star is the bright source in the centre of the map.}
\label{fig:GammaDra}      
\end{figure}
% ***************************************************************************

\begin{figure*}
\centering
  \includegraphics[width=1.0\textwidth]{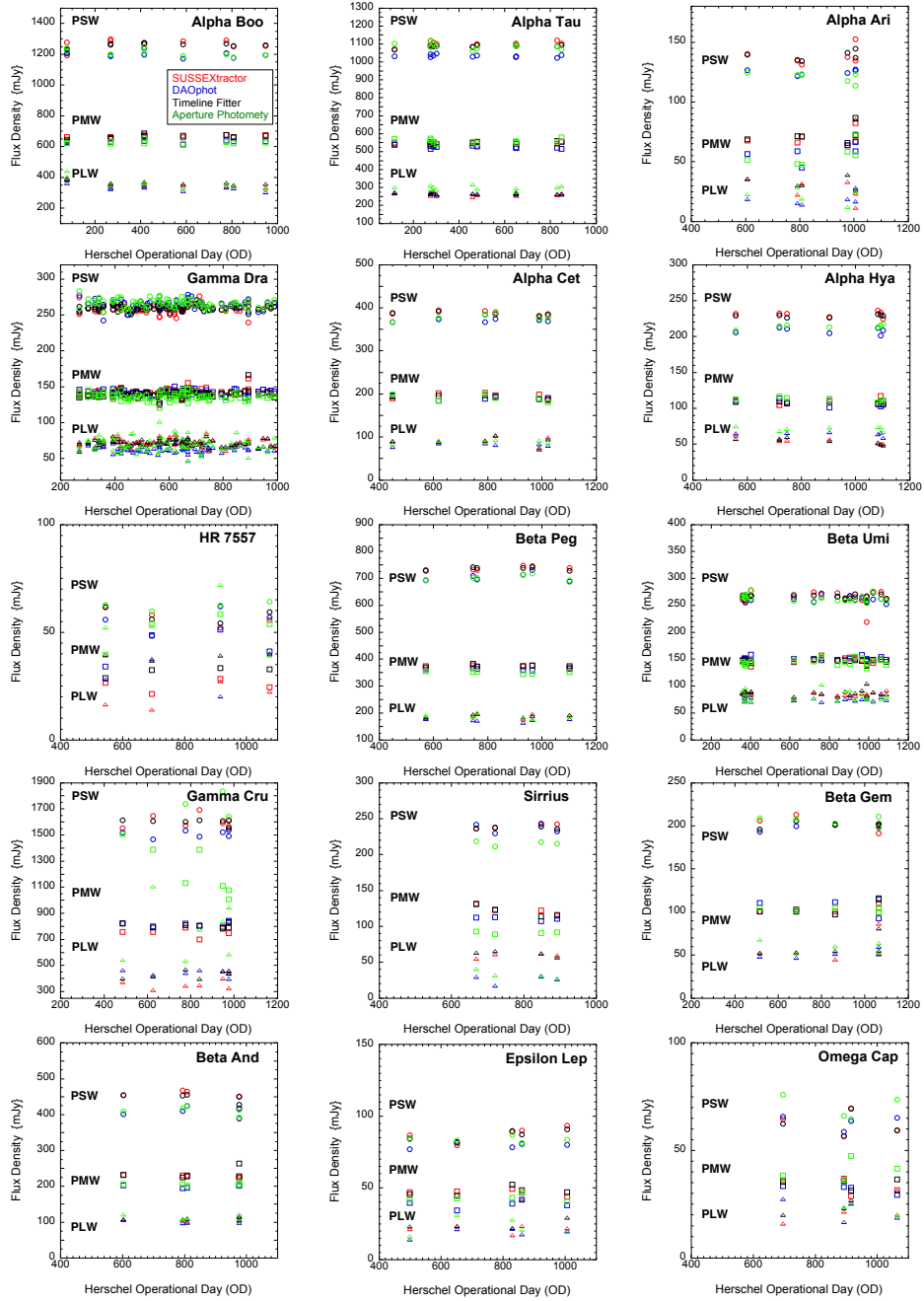}
\caption{Results of the photometry on the SPIRE standard calibration stars. Each panel is for a different calibration star with results for the SPIRE PSW (250$\mu$m), PMW (350$\mu$m), PLW (500$\mu$m)  bands shown as decreasing flux densities (circles, squares and triangles respectively). Results are shown for SUSSEXtractor (red), DAOphot (blue), Timeline Fitter (black), Aperture Photometry (green).}
\label{fig:calibrationstars}       
\end{figure*}

The results for individual stars are shown in Figure~\ref{fig:calibrationstars} with measured flux density plotted against Operational Day for each star, for each SPIRE band. From the figure it can be seen that the photometry methods are in broad general agreement, however variations, offsets and scatter is also apparent. 

The star most often observed for calibration purposes with SPIRE is Gamma Draconis (with example final maps shown in Figure~\ref{fig:GammaDra}) which is a fairly faint source with mean fluxes of 260mJy, 140mJy \& 73mJy in the  PSW, PMW \& PLW bands respectively as measured by the Timeline Fitter method. The mean flux measured by SUSSEXtractor, DAOphot and the Timeline Fitter methods all  agree well with each other in the PSW band to within $\pm$4mJy except for the Aperture Photometry method that reports an average flux closer to 276mJy. For Gamma  Draconis, the PMW and PLW bands, all photometry methods agree to within 1mJy for the mean flux measurements. However, the dispersion from measurement to measurement does vary depending on the photometry method. Using Gamma Dra as an example although the same trend is seen for all observations, the Timeline Fitter consistently produces the lowest scatter between measurements with a standard deviation of 3.2mJy compared to 5.5-6.5mJy for the other 3 methods. 

From Figure ~\ref{fig:calibrationstars}, we also see that there are some systematic discrepancies between algorithms in the fluxes measured in some cases. For example, there does seem to be a trend for DAOphot to underestimate the flux in the PSW band by between a few to up to 8$\%$. This is most prominently shown in the case of Alpha Boo, Beta And, Alpha Hydra and Alpha Ari and in many cases the Aperture Photometry also shows a similar discrepancy. For the case of Alpha Boo, further analysis was carried out by varying the background annulus and the aperture radius but neither significantly affected the measured flux. It is possible that the aperture correction is not optimazed in the PSW band.

In Tables~\ref{Tab:StandardstarPhotometryPSW},~\ref{Tab:StandardstarPhotometryPMW},~\ref{Tab:StandardstarPhotometryPLW} the mean fluxes and the percentage standard deviation measured for all stars for all photometry methods is shown for the PSW, PMW \& PLW bands respectively. For the PSW band, we find for all photometry methods, in almost all cases for sources brighter than 100mJy percentage standard deviations of $<$1-2$\%$. For the PMW band, the  percentage standard deviation is still usually $<$3-4$\%$ in most cases. The PLW band shows a lot more scatter but  for the Timeline Fitter is still generally $<$3$\%$. The final mean measurements and percentage standard deviation for all the photometry methods, for all calibration stars, are summarised in Figures~\ref{fig:repeatabilityPSW},~\ref{fig:repeatabilityPMW} \&~\ref{fig:repeatabilityPLW} for the SPIRE PSW, PMW \& PLW bands respectively.

In Tables~\ref{Tab:StandardstarPhotometryPSWmodel},~\ref{Tab:StandardstarPhotometryPMWmodel},~\ref{Tab:StandardstarPhotometryPLWmodel} , a comparison is made between the flux density measured by   the four photometry algorithms within HIPE with the photospheric models derived from the stellar models of \citet{cohen03} and \citet{decin03} (See Lim et al. 2013, in preparation, for details of these models adopted for the SPIRE instrument calibration). In Tables~\ref{Tab:StandardstarPhotometryPSWmodel},~\ref{Tab:StandardstarPhotometryPMWmodel},~\ref{Tab:StandardstarPhotometryPLWmodel} the mean deviation of the model from each photometry method and the scatter (standard deviation) of each ensemble of observations is tabulated. A comparison between the HIPE photometry and the models shows that in most cases there is agreement in the deviation from the mean measured/model ratio at the $<$10$\%$ level although there are notable exceptions, e.g. Gamma Cru and Beta Gem. The scatter reflects the results of Tables~\ref{Tab:StandardstarPhotometryPSW},~\ref{Tab:StandardstarPhotometryPMW},~\ref{Tab:StandardstarPhotometryPLW}, with the Timeline Fitter producing the most consistent results. The models used for comparison here do not include any dust excess at the SPIRE wavelengths and hence should not be considered as the true flux density of the calibration stars.  The excess emission due to long wavelength dust and a more thorough comparison with available models will be addressed in  Lim et al. 2013 (in preparation).

%
% ************************* BEGIN FIGURE *************************
% ***************************************************************************

%   *************************  BEGIN TABLE  *************************
\begin{table*}
\caption{Results for photometry made in the SPIRE PSW (250$\mu$m) band on the SPIRE standard calibration stars for the four photometry algorithms within HIPE. The mean flux from each ensemble of observations is tabulated along with the percentage standard deviation.}
\label{Tab:StandardstarPhotometryPSW}       % Give a unique label
\begin{tabular}{lllllllll}
\hline\noalign{\smallskip}    
Star	&	\multicolumn{2}{c}{SUSSEXtractor} 	&	\multicolumn{2}{c}{DAOphot} 	&	\multicolumn{2}{c}{Timeline Fitter}  	&	\multicolumn{2}{c}{Aperture Photometry} \\
   & mean &	 $\%$STD &mean&	 $\%$STD&mean&	 $\%$STD	&mean&	 $\%$STD\\
            & mJy  &	 mJy &mJy&	 mJy&mJy&	 mJy	&mJy&	mJy\\
\noalign{\smallskip}\hline\noalign{\smallskip}
Alpha Boo	&	1275	&	1.23	&	1206	&	1.94	&	1265	&	0.51	&	1259	&	1.92	\\
Alpha Tau 	&	1099	&	1.31	&	1038	&	0.97	&	1087	&	0.65	&	1136	&	1.62	\\
Alpha Ari	&	136	&	2.96	&	125	&	1.75	&	138	&	2.06	&	125	&	2.82	\\
Gamma Dra 	&	259	&	2.08	&	263	&	2.41	&	260	&	1.25	&	276	&	2.37	\\
Alpha Cet	&	389	&	1.32	&	371	&	1.08	&	386	&	1.08	&	392	&	1.36	\\
Alpha Hya	&	231	&	1.50	&	206	&	1.98	&	229	&	0.88	&	220	&	1.36	\\
HR 7557	&	57	&	9.50	&	56	&	8.46	&	58	&	5.86	&	65	&	3.92	\\
Beta Peg	&	740	&	1.59	&	714	&	1.93	&	736	&	1.05	&	747	&	2.20	\\
Beta Umi 	&	265	&	2.25	&	263	&	1.97	&	266	&	1.30	&	273	&	1.93	\\
Gamma Cru	&	1612	&	2.89	&	1506	&	1.64	&	1608	&	0.30	&	1771	&	4.40	\\
Sirius	&	241	&	2.83	&	239	&	3.51	&	238	&	1.40	&	228	&	5.77	\\
Beta Gem 	&	205	&	3.07	&	200	&	1.97	&	201	&	2.04	&	216	&	2.01	\\
Beta And	&	455	&	1.81	&	410	&	1.67	&	453	&	0.43	&	426	&	1.24	\\
Epsilon Lep	&	87	&	6.94	&	80	&	2.60	&	87	&	4.61	&	86	&	4.43	\\
Omega Cap 	&	60	&	11.04	&	64	&	4.87	&	61	&	8.41	&	74	&	7.42	\\
\noalign{\smallskip}\hline
\end{tabular}
\end{table*}
%   *************************  END TABLE  *************************

%   *************************  BEGIN TABLE  *************************
\begin{table*}
\caption{Results for photometry made in the SPIRE PMW (350$\mu$m) band on the SPIRE standard calibration stars for the four photometry algorithms within HIPE. The mean flux from each ensemble of observations is tabulated along with the percentage standard deviation.}
\label{Tab:StandardstarPhotometryPMW}       % Give a unique label
\begin{tabular}{lllllllll}
\hline\noalign{\smallskip}    
Star	&	\multicolumn{2}{c}{SUSSEXtractor} 	&	\multicolumn{2}{c}{DAOphot} 	&	\multicolumn{2}{c}{Timeline Fitter}  	&	\multicolumn{2}{c}{Aperture Photometry} \\
   & mean &	 $\%$STD &mean&	 $\%$STD&mean&	 $\%$STD	&mean&	 $\%$STD\\
            & mJy  &	 mJy &mJy&	 mJy&mJy&	 mJy	&mJy&	mJy\\
\noalign{\smallskip}\hline\noalign{\smallskip}
Alpha Boo 	&	665	&	1.26	&	627	&	2.79	&	667	&	1.61	&	650	&	2.78	\\
Alpha Tau 	&	541	&	1.38	&	524	&	1.25	&	551	&	0.97	&	568	&	1.97	\\
Alpha Ari	&	67	&	3.96	&	56	&	13.39	&	70	&	3.82	&	53	&	7.56	\\
Gamma Dra 	&	141	&	2.67	&	139	&	2.99	&	140	&	2.23	&	141	&	4.40	\\
Alpha Cet 	&	194	&	3.33	&	186	&	2.48	&	193	&	1.74	&	196	&	4.72	\\
Alpha Hya	&	108	&	3.59	&	104	&	3.79	&	109	&	1.79	&	113	&	4.08	\\
HR 7557 	&	25	&	14.67	&	44	&	15.86	&	33	&	6.40	&	55	&	10.59	\\
Beta Peg 	&	373	&	0.79	&	360	&	1.48	&	375	&	0.90	&	368	&	2.76	\\
Beta Umi	&	145	&	3.00	&	147	&	2.89	&	147	&	1.42	&	148	&	4.48	\\
Gamma Cru	&	763	&	3.77	&	773	&	1.90	&	801	&	1.65	&	934	&	7.31	\\
Sirius	&	121	&	3.91	&	113	&	5.91	&	120	&	4.35	&	100	&	10.40	\\
Beta Gem	&	102	&	2.77	&	103	&	9.27	&	99	&	1.70	&	109	&	5.45	\\
Beta And	&	226	&	2.17	&	193	&	0.75	&	229	&	1.15	&	204	&	0.86	\\
Epsilon Lep	&	47	&	3.86	&	37	&	4.40	&	48	&	7.07	&	41	&	7.83	\\
Omega Cap	&	35	&	11.98	&	31	&	8.88	&	37	&	17.49	&	42	&	20.73	\\
\noalign{\smallskip}\hline
\end{tabular}
\end{table*}
%   *************************  END TABLE  *************************

%   *************************  BEGIN TABLE  *************************
\begin{table*}
\caption{Results for photometry made in the SPIRE PLW (500$\mu$m) band on the SPIRE standard calibration stars for the four photometry algorithms within HIPE. The mean flux from each ensemble of observations is tabulated along with the percentage standard deviation.}
\label{Tab:StandardstarPhotometryPLW}       % Give a unique label
\begin{tabular}{lllllllll}
\hline\noalign{\smallskip}    
Star	&	\multicolumn{2}{c}{SUSSEXtractor} 	&	\multicolumn{2}{c}{DAOphot} 	&	\multicolumn{2}{c}{Timeline Fitter}  	&	\multicolumn{2}{c}{Aperture Photometry} \\
   & mean &	 $\%$STD &mean&	 $\%$STD&mean&	 $\%$STD	&mean&	 $\%$STD\\
         & mJy  &	 mJy &mJy&	 mJy&mJy&	 mJy	&mJy&	mJy\\
\noalign{\smallskip}\hline\noalign{\smallskip}
Alpha Boo 	&	348	&	3.05		&	319	&	4.48		&	360	&	1.76		&	352	&	4.56	\\
Alpha Tau 	&	259	&	2.96		&	265	&	2.70		&	268	&	1.82		&	303	&	4.11	\\
Alpha Ari		&	28	&	25.01	&	16	&	18.48	&	32	&	16.15	&	19	&	26.22	\\
Gamma Dra 	&	71	&	6.83		&	62	&	7.83		&	73	&	5.10		&	71	&	12.10	\\
Alpha Cet 	&	90	&	12.09	&	79	&	4.48		&	91	&	10.22	&	98	&	12.15	\\
Alpha Hya		&	53	&	10.19	&	62	&	6.60		&	55	&	7.12		&	78	&	15.75	\\
HR 7557 		&	20	&	31.30	&	31	&	32.96	&	39	&	3.34		&	50	&	27.19	\\
Beta Peg 		&	187	&	3.48		&	170	&	5.56		&	190	&	2.85		&	189	&	7.56	\\
Beta Umi		&	84	&	4.45		&	74	&	6.74		&	87	&	3.90		&	83	&	11.44	\\
Gamma Cru	&	362	&	7.78		&	426	&	3.73		&	431	&	7.12		&	514	&	7.19	\\
Sirius		&	56	&	5.76		&	35	&	26.16	&	62	&	5.56		&	36	&	31.62	\\
Beta Gem		&	50	&	6.81		&	48	&	7.47		&	54	&	2.29		&	60	&	16.11	\\
Beta And		&	105	&	2.36		&	100	&	7.78		&	110	&	2.83		&	113	&	4.43	\\
Epsilon Lep	&	21	&	11.76	&	17	&	11.97	&	28	&	27.15	&	20	&	24.89	\\
Omega Cap	&	20	&	14.96	&	22	&	23.40	&	25	&	16.21	&	28	&	18.62	\\
\noalign{\smallskip}\hline
\end{tabular}
\end{table*}
%   *************************  END TABLE  *************************

%   *************************  BEGIN TABLE  *************************
\begin{table*}
\caption{Comparison of photospheric models with photometry made in the SPIRE PSW (250$\mu$m) band on the SPIRE standard calibration stars for the four photometry algorithms within HIPE. The mean deviation from the model and scatter (standard deviation) of each ensemble of observations is tabulated.}
\label{Tab:StandardstarPhotometryPSWmodel}       % Give a unique label
\begin{tabular}{llllllllll}
\hline\noalign{\smallskip}    
Star	& model &	\multicolumn{2}{c}{SUSSEXtractor} 	&	\multicolumn{2}{c}{DAOphot} 	&	\multicolumn{2}{c}{Timeline Fitter}  	&	\multicolumn{2}{c}{Aperture Phot.} \\
   &  & mean  &	 STD & mean &	 STD & mean &	 STD	& mean &	 STD\\
      & mJy & mJy  &	 mJy & mJy &	 mJy & mJy &	 mJy	& mJy &	mJy\\
\noalign{\smallskip}\hline\noalign{\smallskip}
Alpha Boo 	&1163$^{D}$	&	1.09	&	0.03	&	1.03	&	0.02	&	1.08		&	0.01&	1.04	&	0.02	\\
Alpha Tau 	&1083$^{D}$	&	1.02	&	0.01	&	0.95	&	0.01	&	1.00		&	0.01&	1.01	&	0.01	\\
Alpha Ari		&126$^{C}$	&	1.10	&	0.06	&	0.99	&	0.02	&	1.10		&	0.03&	0.96	&	0.03	\\
Gamma Dra 	&252$^{D}$	&	1.02	&	0.07	&	1.04	&	0.03	&	1.03		&	0.03&	1.06	&	0.03	\\
Alpha Cet 	&375$^{D}$	&	1.03	&	0.01	&	0.99	&	0.01	&	1.03		&	0.01&	1.01	&	0.02	\\
Alpha Hya		&211$^{C}$	&	1.09	&	0.02	&	0.99	&	0.02	&	1.08		&	0.01&	1.01	&	0.01	\\
HR 7557 		&-		&	-	&	-	&	-	&	-	&	-	&	-	&	-	&	-	\\
Beta Peg 		&662$^{D}$	&	1.11	&	0.02	&	1.07	&	0.02	&	1.11		&	0.01&	1.07	&	0.02	\\
Beta Umi		&264$^{C}$	&	1.00	&	0.05	&	0.99	&	0.02	&	1.01		&	0.02&	1.00	&	0.02	\\
Gamma Cru	&1397$^{C}$	&	1.15	&	0.04	&	1.08	&	0.02	&	1.15		&	0.02&	1.38	&	0.22	\\
Sirius		&219$^{D}$	&	1.10	&	0.03	&	1.09	&	0.02	&	1.09		&	0.02&	1.00	&	0.02	\\
Beta Gem		&179$^{C}$	&	1.13	&	0.04	&	1.11	&	0.02	&	1.12		&	0.02&	1.15	&	0.03	\\
Beta And		&432$^{D}$	&	1.04	&	0.05	&	0.94	&	0.03	&	1.04		&	0.03&	0.95	&	0.03	\\
Epsilon Lep	&100$^{C}$	&	0.88	&	0.05	&	0.79	&	0.02	&	0.87		&	0.04&	0.84	&	0.02	\\
Omega Cap	&-		&	-	&	-	&	-	&	-	&	-	&	-	&	-	&	-	\\
\noalign{\smallskip}\hline
\multicolumn{10}{l}{	$D$: Photospheric models of  \citet{decin03}}  \\
\multicolumn{10}{l}{	$C$: Photospheric models of  \citet{cohen03}}  \\
\end{tabular}
\end{table*}
%   *************************  END TABLE  *************************

%   *************************  BEGIN TABLE  *************************
\begin{table*}
\caption{Comparison of photospheric models with photometry made in the SPIRE PMW (350$\mu$m) band on the SPIRE standard calibration stars for the four photometry algorithms within HIPE. The mean deviation from the model and scatter (standard deviation) of each ensemble of observations is tabulated.}
\label{Tab:StandardstarPhotometryPMWmodel}       % Give a unique label
\begin{tabular}{llllllllll}
\hline\noalign{\smallskip}    
Star	& model &	\multicolumn{2}{c}{SUSSEXtractor} 	&	\multicolumn{2}{c}{DAOphot} 	&	\multicolumn{2}{c}{Timeline Fitter}  	&	\multicolumn{2}{c}{Aperture Phot.} \\
   &  & mean  &	 STD & mean &	 STD & mean &	 STD	& mean &	 STD\\
      & mJy & mJy  &	 mJy & mJy &	 mJy & mJy &	 mJy	& mJy &	mJy\\
\noalign{\smallskip}\hline\noalign{\smallskip}
Alpha Boo 	&584$^{D}$	&	1.09	&	0.02	&	1.14	&	0.02	&	1.07	&	0.02	&	1.27	&	0.02	\\
Alpha Tau 	&546$^{D}$	&	1.00	&	0.02	&	0.96	&	0.01	&	1.01	&	0.01	&	1.02	&	0.02	\\
Alpha Ari		&64$^{C}$	&	1.09	&	0.10	&	0.91	&	0.12	&	1.13	&	0.11	&	0.87	&	0.15	\\
Gamma Dra 	&128$^{D}$	&	1.11	&	0.04	&	1.11	&	0.03	&	1.10	&	0.04	&	1.07	&	0.04	\\
Alpha Cet 	&189$^{D}$	&	1.04	&	0.04	&	1.01	&	0.02	&	1.02	&	0.02	&	1.01	&	0.04	\\
Alpha Hya		&211$^{C}$	&	1.04	&	0.04	&	1.00	&	0.04	&	1.02	&	0.02	&	1.06	&	0.02	\\
HR 7557 		&-		&	-	&	-	&	-	&	-	&	-	&	-	&	-	&	-	\\
Beta Peg 		&334$^{D}$	&	1.12	&	0.01	&	1.09	&	0.02	&	1.12	&	0.01	&	1.06	&	0.03	\\
Beta Umi		&133$^{C}$	&	1.10	&	0.04	&	1.13	&	0.03	&	1.10	&	0.02	&	1.08	&	0.04	\\
Gamma Cru	&704$^{C}$	&	1.09	&	0.06	&	1.16	&	0.02	&	1.14	&	0.02	&	1.75	&	0.27	\\
Sirius		&110$^{D}$	&	1.12	&	0.04	&	1.01	&	0.05	&	1.10	&	0.05	&	0.82	&	0.03	\\
Beta Gem		&90$^{C}$	&	1.15	&	0.04	&	1.18	&	0.10	&	1.14	&	0.08	&	1.15	&	0.03	\\
Beta And		&219$^{D}$	&	1.04	&	0.02	&	0.91	&	0.02	&	1.08	&	0.07	&	0.94	&	0.02	\\
Epsilon Lep	&50$^{C}$	&	0.94	&	0.04	&	0.77	&	0.05	&	0.95	&	0.06	&	0.86	&	0.05	\\
Omega Cap	&-		&	-	&	-	&	-	&	-	&	-	&	-	&	-	&	-	\\
\noalign{\smallskip}\hline
\multicolumn{10}{l}{	$D$: Photospheric models of  \citet{decin03}}  \\
\multicolumn{10}{l}{	$C$: Photospheric models of  \citet{cohen03}}  \\
\end{tabular}
\end{table*}
%   *************************  END TABLE  *************************

%   *************************  BEGIN TABLE  *************************
\begin{table*}
\caption{Comparison of photospheric models with photometry made in the SPIRE PLW (500$\mu$m) band on the SPIRE standard calibration stars for the four photometry algorithms within HIPE. The mean deviation from the model and scatter (standard deviation) of each ensemble of observations is tabulated.}
\label{Tab:StandardstarPhotometryPLWmodel}       % Give a unique label
\begin{tabular}{llllllllll}
\hline\noalign{\smallskip}    
Star	& model &	\multicolumn{2}{c}{SUSSEXtractor} 	&	\multicolumn{2}{c}{DAOphot} 	&	\multicolumn{2}{c}{Timeline Fitter}  	&	\multicolumn{2}{c}{Aperture Phot.} \\
   &  & mean  &	 STD & mean &	 STD & mean &	 STD	& mean &	 STD\\
      & mJy & mJy  &	 mJy & mJy &	 mJy & mJy &	 mJy	& mJy &	mJy\\
\noalign{\smallskip}\hline\noalign{\smallskip}
Alpha Boo 	&281$^{D}$	&	1.27	&	0.08	&	1.20	&	0.09	&	1.30	&	0.04	&	1.29	&	0.12	\\
Alpha Tau 	&264$^{D}$	&	0.99	&	0.03	&	1.01	&	0.02	&	1.01	&	0.02	&	1.14	&	0.05	\\
Alpha Ari		&31$^{C}$	&	0.85	&	0.29	&	0.61	&	0.16	&	1.33	&	0.71	&	0.72	&	0.19	\\
Gamma Dra 	&62$^{D}$	&	0.85	&	0.29	&	0.61	&	0.16	&	1.33	&	0.71	&	0.72	&	0.19	\\
Alpha Cet 	&92$^{D}$	&	0.99	&	0.12	&	0.90	&	0.04	&	0.99	&	0.10	&	0.98	&	0.04	\\
Alpha Hya		&51$^{C}$	&	1.06	&	0.09	&	1.24	&	0.08	&	1.07	&	0.08	&	1.41	&	0.06	\\
HR 7557 		&-		&	-	&	-	&	-	&	-	&	-	&	-	&	-	&	-	\\
Beta Peg 		&161$^{D}$	&	1.17	&	0.04	&	1.06	&	0.06	&	1.18	&	0.03	&	1.15	&	0.08	\\
Beta Umi		&64$^{C}$	&	1.33	&	0.06	&	1.19	&	0.07	&	1.37	&	0.08	&	1.26	&	0.14	\\
Gamma Cru	&339$^{C}$	&	1.08	&	0.15	&	1.32	&	0.07	&	1.27	&	0.08	&	2.25	&	0.65	\\
Sirius		&53$^{D}$	&	1.10	&	0.08	&	0.50	&	0.08	&	1.18	&	0.07	&	0.60	&	0.08	\\
Beta Gem		&44$^{C}$	&	1.31	&	0.37	&	1.17	&	0.12	&	1.35	&	0.28	&	1.35	&	0.16	\\
Beta And		&107$^{D}$	&	1.00	&	0.05	&	0.97	&	0.08	&	1.02	&	0.03	&	1.07	&	0.06	\\\
Epsilon Lep	&24$^{C}$	&	0.89	&	0.11	&	0.79	&	0.13	&	1.17	&	0.34	&	0.97	&	0.26	\\
Omega Cap	&-		&	-	&	-	&	-	&	-	&	-	&	-	&	-	&	-	\\
\noalign{\smallskip}\hline
\multicolumn{10}{l}{	$D$: Photospheric models of  \citet{decin03}}  \\
\multicolumn{10}{l}{	$C$: Photospheric models of  \citet{cohen03}}  \\
\end{tabular}
\end{table*}
%   *************************  END TABLE  *************************

\subsection{Dark Skies}
\label{sec:darkskies}
\begin{figure}
\centering
  \includegraphics[width=0.75\textwidth]{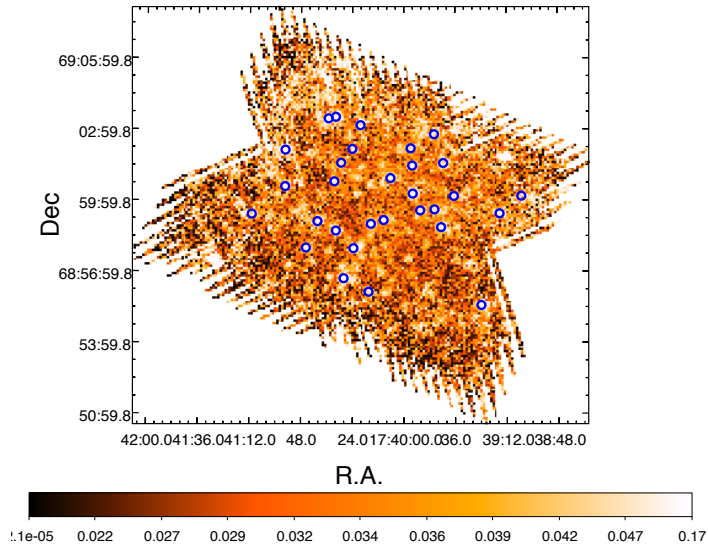}
  \caption{PSW (250$\mu$m) image of the SPIRE dark sky observation used to create the initial source list for the photometry analysis. Extracted sources are shown as blue circles.}
\label{fig:darksky}      
\end{figure}

The SPIRE standard calibration stars cover the flux range from sources brighter than 60mJy at 250$\mu$m to sources brighter than a Jansky.  In order to probe the flux density range to levels down to 10mJy, fainter sources are desirable. Since there are no standard stars for SPIRE this faint, we instead turn to the SPIRE dark sky observations. The SPIRE dark sky is an area of sky centred near the North Ecliptic Pole at R.A. = 17h40m12s, Dec = +69d00m00s and was observed on many times as part of the SPIRE routine calibration plan. We  selected 70 dark sky observations taken between Operational Days 300 -- 1156 (11th March 2011 -- 12th July 2012) for our photometry measurements. All observations were processed using the standard SPIRE Small Map pipeline using the HIPE 11 SPIRE Calibration Tree. We then carried out source extraction with SUSSEXractor on a single dark sky image (observation ID = 1342247979) with a detection threshold = 5, to retrieve a total of 40 sources. The source positions were visually inspected and any sources lying at the periphery of the map, where the coverage was low, were discarded leaving a total of 34 remaining sources. The sources were colour corrected assuming a Rayleigh-Jeans approximation of $F_{\nu}\propto \nu^{\alpha}$, $ \alpha$=2 which by inspection of the measured fluxes in the 3 SPIRE bands, appears reasonable for the majority of our sources. The dark sky observation used for the initial source extraction is shown in  Figure~\ref{fig:darksky}. The extracted sources range from 60--20 mJy, 40--10 mJy and 30--3 mJy in the PSW, PMW and PLW bands respectively. Note that the SPIRE confusion limit as measured by ~\citet{nguyen10} is 5.8, 6.3 and 6.8 mJy in the short to long wavelength SPIRE bands, which means that the faintest objects we measure in the PLW (and probably PMW) bands may be severely affected by confusion due to unresolved sources.

The  positions for the extracted sources were then used as the input parameters for the other photometry algorithms. The same source list was used for all of the 70 dark sky images. Note that the variation of the roll angle of the spacecraft with observation date means that different observations can have different orientations (in fact for SPIRE small maps, only the central 5 arcmins is guaranteed for scientific observations ), therefore, the coverage map value at each source position was also examined and if the source position was either off the map area or in areas of low coverage, the measurement of that particular source for that particular observation was not included in the final analysis.

% ************************* BEGIN FIGURE *************************
% ***************************************************************************

For the brighter ($F_{250\mu m}\sim$50mJy) sources in our dark sky sample, the results agree well with the calibration stars of similar flux density (e.g. Alpha Ari, Epsilon Lep), with the Timeline Fitter providing measurements with less scatter than the other photometry methods with percentage standard deviations of the order of 5$\%$. However, at fluxes fainter than approximately 30 mJy in all bands, the Timeline Fitter begins to struggle and in some cases is unable to provide a fit. This is not suprising given the faint level of these sources. At flux levels $<$30mJy the percentage standard deviation on the measurement of the source flux densities has risen to of the order of 30$\%$ or more with SUSSEXtractor (and also the Aperture Photometry) providing better estimates than DAOphot and the Timeline Fitter. The final  summary of the repeatability of photometry on the dark sky observations is shown in Figures~\ref{fig:repeatabilityPSW},\ref{fig:repeatabilityPMW} \& \ref{fig:repeatabilityPLW} for the SPIRE PSW, PMW \& PLW bands respectively.

% *************************************************************************************************
% ***********                        Section: Summary             *********************************
% *************************************************************************************************
\section{Summary and Conclusions}
\label{sec:summary}
\begin{figure}
\centering
  \includegraphics[width=0.6\textwidth]{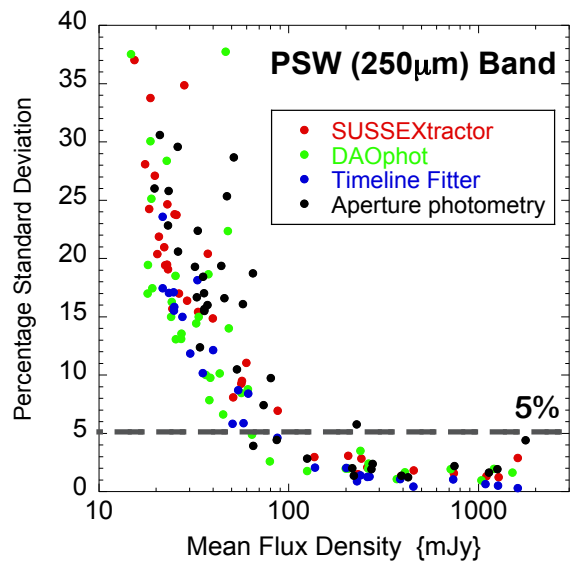}
\caption{Repeatability of photometry for the photometry methods available in HIPE for  the PSW (250$\mu$m) band. Percentage standard deviation from the mean is shown against mean flux density for all calibration stars and SPIRE dark sky observations. }
\label{fig:repeatabilityPSW}      
\end{figure}

\begin{figure}
\centering
  \includegraphics[width=0.6\textwidth]{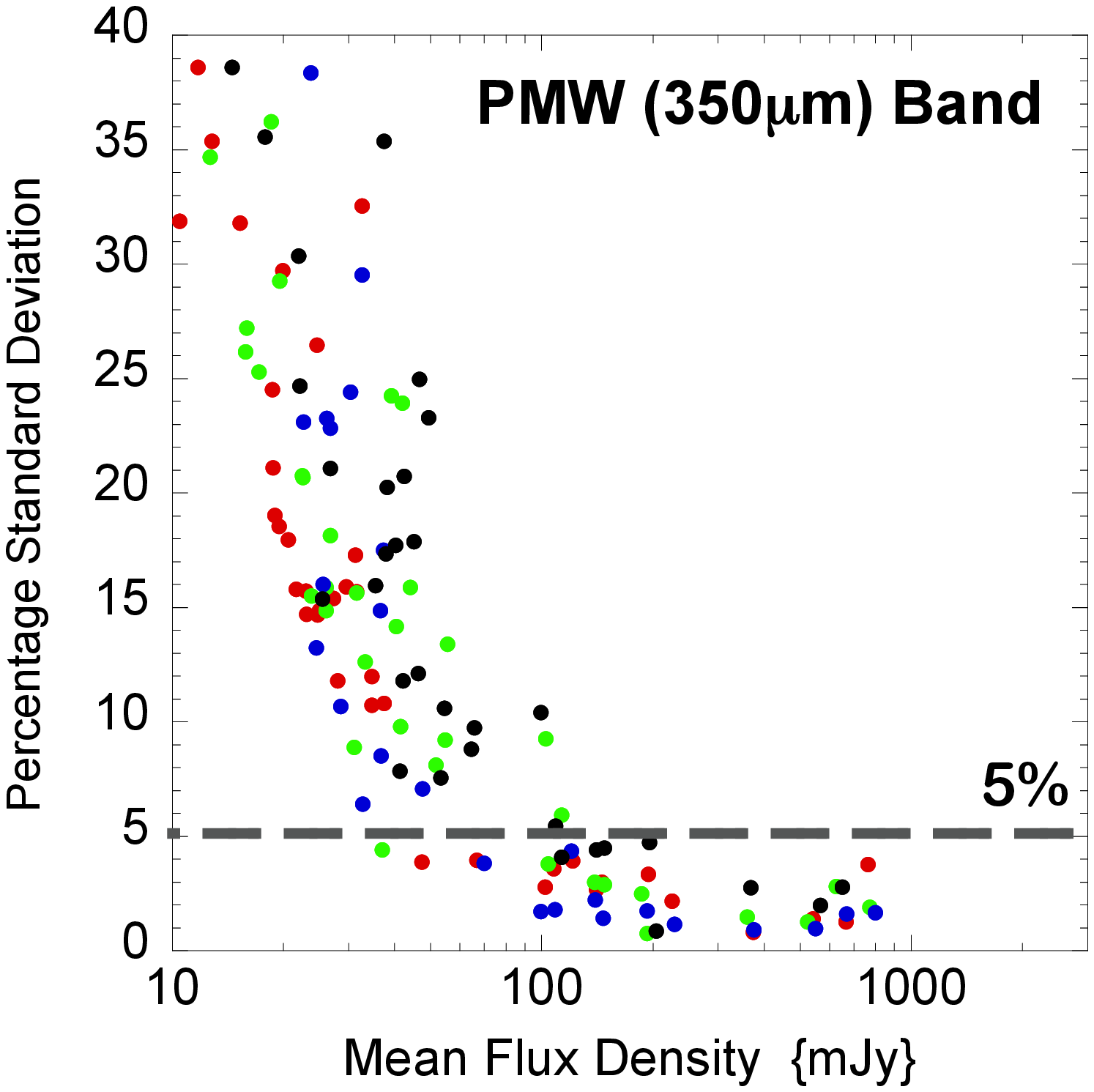}
\caption{Repeatability of photometry for the photometry methods available in HIPE for  the PMW (350$\mu$m) band. Percentage standard deviation from the mean is shown against mean flux density for all calibration stars and SPIRE dark sky observations. }
\label{fig:repeatabilityPMW}      
\end{figure}

\begin{figure}
\centering
  \includegraphics[width=0.6\textwidth]{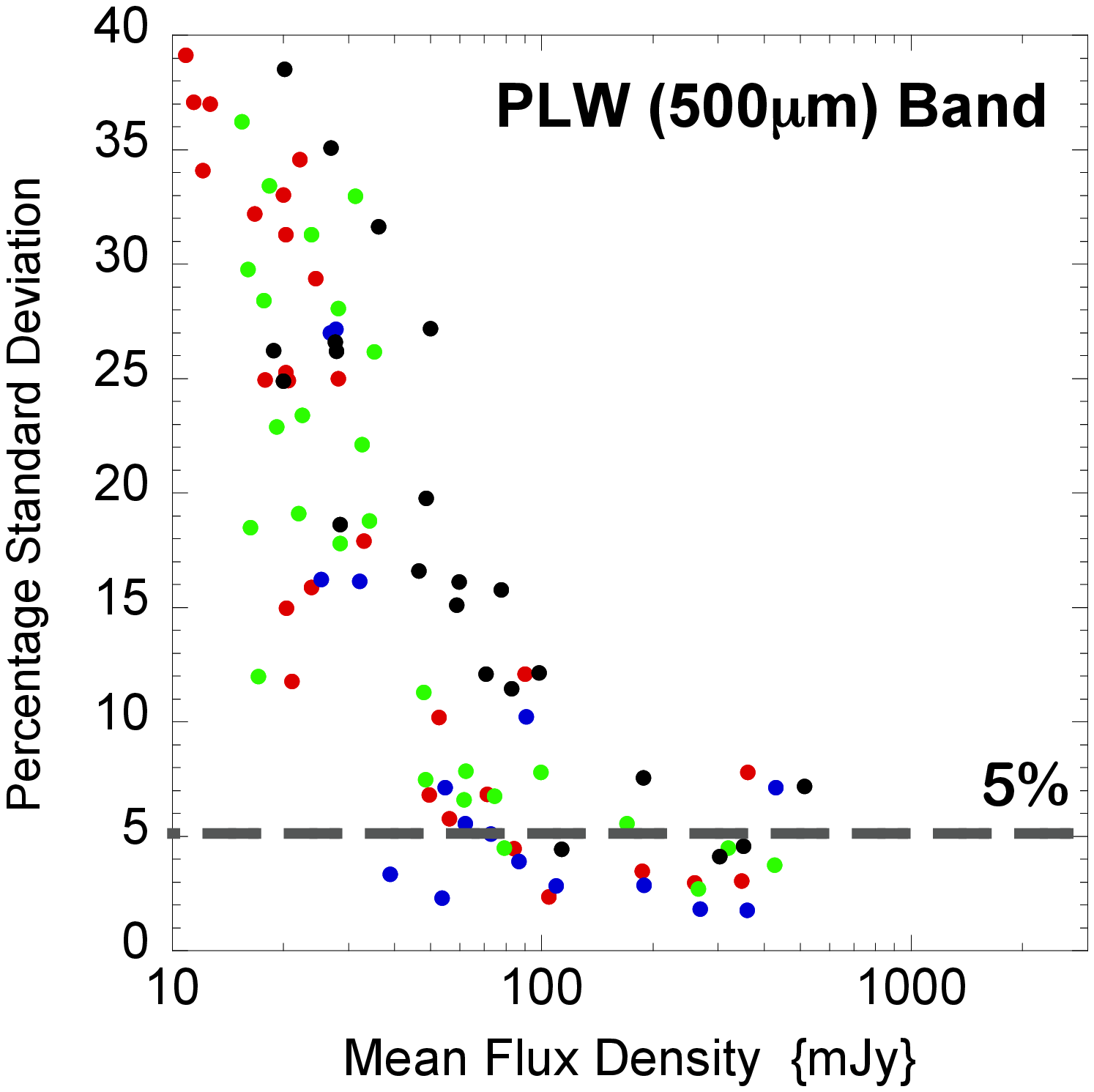}
\caption{Repeatability of photometry for the photometry methods available in HIPE for  the PLW (500$\mu$m) band. Percentage standard deviation from the mean is shown against mean flux density for all calibration stars and SPIRE dark sky observations. }
\label{fig:repeatabilityPLW}      
\end{figure}

In Figures ~\ref{fig:repeatabilityPSW},\ref{fig:repeatabilityPMW} \& \ref{fig:repeatabilityPLW}, the results for the photometry measurements on the SPIRE standard calibration stars and the SPIRE Dark Sky observations are summarised by plotting the measured flux density and the percentage standard deviation. The percentage standard deviation provides a measure of the repeatability of any given photometry method.

% ************************* BEGIN FIGURE *************************
% ***************************************************************************

% ************************* BEGIN FIGURE *************************
% ***************************************************************************

% ************************* BEGIN FIGURE *************************
% ***************************************************************************

For the PSW band, from Figure~\ref{fig:repeatabilityPSW}  it can be seen that down to approximately 100mJy, the reproducibility  of photometric measurements is of the order of 2-3$\%$ for all the photometry algorithms that were tested. Moreover, the Timeline Fitter provides reliable, repeatable measurements at the 1-2$\%$ level in this flux range. In the intermediate flux range 30--150mJy, DAOphot is also impressive in terms of repeatability, however in many cases underestimates the source flux by $\sim$8$\%$. At the faintest fluxes, SUSSEXtractor performs well. 

For the PMW band a similar trend is seen, where the Timeline Fitter produces consistent photometry to the level of a few percent down to the 100mJy level. All photometry methods show a larger scatter at fainter fluxes with again SUSSEXtractor providing slightly better reliability as the fluxes move toward the confusion limit.

In the PLW band, the Timeline Fitter again produces excellent repeatability down to fluxes of 100mJy and to considerably fainter fluxes. The Aperture Photometry performs particularly poorly in the PLW band with a large scatter even at brighter fluxes.

\begin{figure*}
\centering
  \includegraphics[width=0.5\textwidth, angle=90]{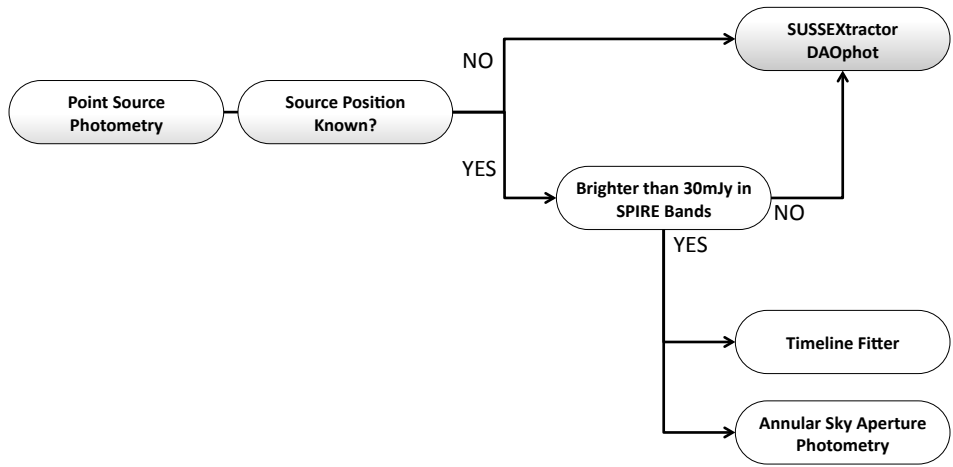}
\caption{Summary organogram for selecting point source photometry algorithms within HIPE.}
\label{fig:photometryflowchart}       
\end{figure*}

In conclusion, we find that the SPIRE Timeline Fitter is the method of choice for source photometry down to the $\sim$30mJy level in all SPIRE bands. Even though at intermediate fluxes, DAOphot produces slightly better repeatability, there exists the possibility of a systematic offset in the measured source fluxes, possibly due to a sub-optimal aperture correction. SUSSEXtractor, although producing a similar average flux as the Timeline Fitter for the total ensemble, suffers from a larger scatter. However, the Timeline Fitter  has trouble fitting the source fluxes at the faintest levels, sometime failing to fit completely. Therefore at flux levels $<$30mJy SUSSEXtractor or DAOphot may provide a better result. The conclusions of the photometry tests made in this work are summarised in Figure~\ref{fig:photometryflowchart}, which shows the recommended route for source extraction and photometry within the HIPE system. 

%
% ************************* BEGIN FIGURE *************************
% ***************************************************************************

\begin{acknowledgements}
The authors would like to thank the referee for providing valuable comments that improved the results of this paper.
SPIRE has been developed by a consortium of institutes led by
Cardiff Univ. (UK) and including: Univ. Lethbridge (Canada);
NAOC (China); CEA, LAM (France); IFSI, Univ. Padua (Italy);
IAC (Spain); Stockholm Observatory (Sweden); Imperial College
London, RAL, UCL-MSSL, UKATC, Univ. Sussex (UK); and
Caltech, JPL, NHSC, Univ. Colorado (USA). This development
has been supported by national funding agencies: CSA (Canada);
NAOC (China); CEA, CNES, CNRS (France); ASI (Italy);MCINN
(Spain); SNSB (Sweden); STFC, UKSA (UK); and NASA (USA).
HIPE is a joint development by the Herschel Science Ground Segment
Consortium, consisting of ESA, the NASA Herschel Science
Center, and the HIFI, PACS and SPIRE consortia.
\end{acknowledgements}



\end{document}